\documentclass[cameraready]{Interspeech}
\def\argmax{\mathop{\rm argmax}}

\title{Self-Speculative Decoding for LLM-based ASR with CTC Encoder Drafts}

\author{George}{Saon}
\author{Samuel}{Thomas}
\author{Takashi}{Fukuda}
\author{Tohru}{Nagano}
\author{Avihu}{Dekel}
\author{Luis}{Lastras}

\address{
    IBM Research
}

\email{gsaon@us.ibm.com}


\newcommand{\blue}[1]{\textcolor{blue}{#1}}

\usepackage{comment}
\usepackage{diagbox}
\usepackage{multirow}
\usepackage{colortbl}

\begin{document}

\maketitle

\begin{abstract}
We propose self-speculative decoding for speech-aware LLMs by using the CTC encoder as a draft model to accelerate auto-regressive (AR) inference and improve ASR accuracy. Our three-step procedure works as follows: (1) if the frame entropies of the CTC output distributions are below a threshold, the greedy CTC hypothesis is accepted as final; (2) otherwise, the CTC hypothesis is verified in a single LLM forward pass using a relaxed acceptance criterion based on token likelihoods; (3) if verification fails, AR decoding resumes from the accepted CTC prefix. Experiments on nine corpora and five languages show that this approach can simultaneously accelerate decoding and reduce WER. On the HuggingFace Open ASR benchmark with a 1B parameter LLM and 440M parameter CTC encoder, we achieve a record 5.58\% WER and improve the inverse real time factor by a factor of 4.4 with only a 12\% relative WER increase over AR search. Code and model weights are publicly available under a permissive license.
    
\end{abstract}

\section{Introduction}

As a special case of attention encoder-decoder (AED) models~\cite{chan2016listen}, speech-aware language models (short SLMs) are currently the best performing ASR systems in terms of recognition accuracy. Indeed, the top five open models on the HuggingFace Open ASR leaderboard~\cite{srivastav2025open} are SLMs with similar architectures. They all use conformer acoustic encoders trained with either RNN-T loss~\cite{rekesh2023fast,sekoyan2025canary}, CTC loss~\cite{saon2025granite}, or with next token cross-entropy loss as part of an AED ASR model~\cite{abouelenin2025phi,shi2026qwen3asrtechnicalreport} and speech modality adapters which are either MLPs~\cite{chen2024salm,abouelenin2025phi} or query transformers~\cite{saon2025granite}. The role of the adapters is to perform a temporal downsampling and to project the acoustic embeddings coming out of the encoder to a space that is interpretable by the LLM. For ASR, the LLM is trained to perform a repetition-like task~\cite{grattafiori2024llama} in the sense that the embeddings of the output tokens can be seen as denoised versions of the speech embeddings. 

One of the main drawbacks of AED models and, by extension, of SLMs is that inference is done auto-regressively, one token at a time, which requires one forward pass through the text LLM per generated token. This limits parallelism compared to non-autoregressive approaches where output tokens are predicted simultaneously such as CTC with greedy decoding, CTC with mask-predict error correction~\cite{higuchi2020mask} or iterative realignment~\cite{chi2021align}. Another class of models which have a good accuracy / inference speed tradeoff, termed semi-autoregressive, predict multiple output tokens at once within blocks of frames~\cite{arora2024semi} or jointly predict tokens and durations~\cite{xu2023efficient}. 

Accelerating AR inference for text LLMs via speculative decoding is an active area of research. In~\cite{leviathan2023fast} and~\cite{chen2023accelerating}, the authors use a small draft AR model that runs in parallel to the large target model and produces $K$ future tokens at a time conditioned on the current hypothesis which get verified by the target model in a single forward pass. If the verification fails, the target model falls back to AR decoding from the first position where the draft and target model distributions differ. The main idea in self speculative decoding is to reuse the target model itself as a draft model either by simultaneously predicting tokens at future positions $i+1\ldots i+N$ with $N$ separate LM heads as in ~\cite{cai2024medusa} or by using predictions from intermediate layers (early exit)~\cite{zhang2024draft}. 

An application of speculative decoding to ASR for transformer-based encoder-decoders is discussed in~\cite{lim2025hybrid} where the authors train a small token and duration transducer (TDT)~\cite{xu2023efficient} decoder together with the frozen encoder which serves as a draft model for the larger causal transformer decoder. There are several differences between~\cite{lim2025hybrid} and our proposed work. First, we reuse the CTC encoder of the SLM instead of training a separate decoder for the draft model. Second, the authors focus on mitigating repetition errors whereas we show that complementary error patterns between CTC and SLM can actually improve WER. Third, we use a confidence metric given by the frame-level entropies of the CTC output distributions~\cite{laptev2023fast} which allows skipping SLM verification entirely for confident CTC hypotheses.

Other techniques that are relevant to our work are two-pass approaches that use a fast model for the first pass such as RNN-T or AR and either rescore candidate hypotheses~\cite{sainath2019two} or refine the output with a slower AR model~\cite{hu2020deliberation}, and joint CTC/attention models with score fusion during inference~\cite{watanabe2017hybrid}. 

As to limitations of our proposed approach, we note that: (1) it requires an SLM with a CTC-trained encoder that is frozen during projector training and LoRA finetuning of the LLM; (2) our technique is limited to ASR and is not applicable to other speech tasks like translation or spoken question answering; (3) the SLM verification of the CTC hypothesis is utterance-based meaning that, if verification fails, the entire utterance has to be AR-decoded from the failed verification point\footnote{While it is possible to dynamically rejoin the CTC draft through local DP, we found this approach to be inefficient for batched inference.} which limits the gains for corpora with low acceptance rates. We mitigate (3) by using a relaxed acceptance criterion which requires the CTC hypothesis to be {\em plausible} under the SLM distribution instead of an exact match of the SLM verification hypothesis.  

The paper is organized as follows: in section~\ref{sec:method} we describe our proposed method, in section~\ref{sec:experiments} we show some experimental evidence of its utility, and in section~\ref{sec:conclusion} we summarize our findings and propose future directions.

\begin{figure*}[t]                                                                                                                                                                       \centering                                                                                                                                                                               \includegraphics[width=\textwidth]{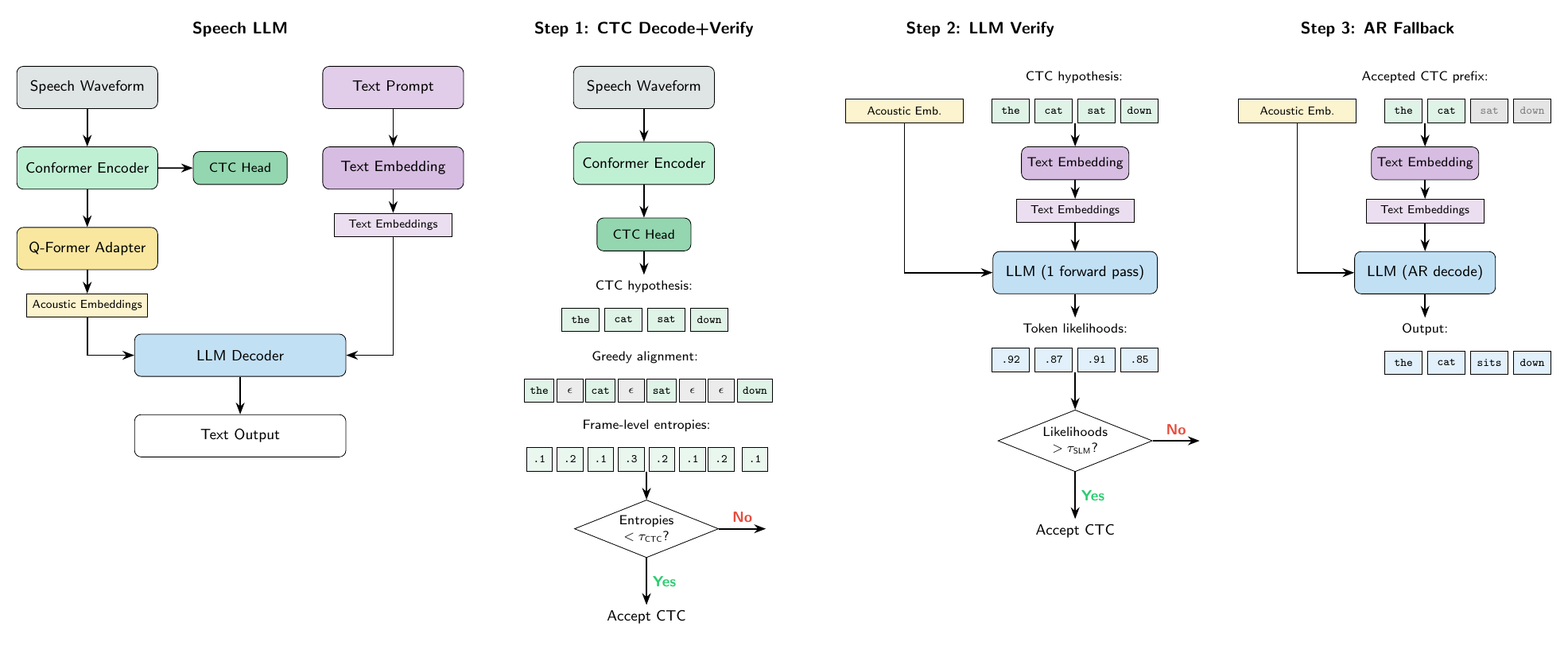}                                                                                                                                   \caption{Overview of the proposed self-speculative decoding method for speech-aware LLMs.}                                                                                               \label{fig:cascade}                                                                                                                                                                      \end{figure*}                                                                                                                                                                                                       
\section{Method description}
\label{sec:method}
In this section we describe our method in detail specifically the CTC decoding and verification pass (subsection~\ref{subsec:CTC}), the SLM verification pass (subsection~\ref{subsec:SLM-verif}), and the SLM auto-regressive fallback pass (subsection~\ref{subsec:SLM-AR}). The three steps as well as the chosen SLM architecture are illustrated in Figure~\ref{fig:cascade}.
\subsection{CTC decoding and verification}
\label{subsec:CTC}
In order to introduce some notations, we recall that the CTC formulation~\cite{graves2006connectionist} models the conditional distribution $p_{CTC}({\bf y}|{\bf x})$ of an output sequence  ${\bf y}=(y_1,\ldots,y_U)\in{\cal Y}^*$ of length $U$ given an input
sequence ${\bf x}=(x_1,\ldots,x_T)\in{\cal X}^*$ of length $T\ge U$. The
elements of ${\bf x}$ are continuous multidimensional
vectors whereas the elements of ${\bf y}$ belong to a discrete output space. $p({\bf y}|{\bf x})$ is expressed as a
sum over all possible alignments ${\bf a}=(a_1,\ldots,a_{T})$ that
are consistent with ${\bf y}$:

\begin{equation}
p_{CTC}({\bf y}|{\bf x})=\sum_{{\bf a}\in{\cal B}^{-1}({\bf y})}p({\bf a}|{\bf x})=\sum_{{\bf a}\in{\cal B}^{-1}({\bf y})}\prod_{t=1}^T p(a_t|h_t)
\label{ll}
\end{equation}
~~\\
where ${\bf h}=(h_1,\ldots,h_T)=Encoder({\bf x})$ is a sequence of acoustic embeddings computed by an encoder and the elements of ${\bf a}$ belong to the augmented vocabulary
$\overline{{\cal Y}}={\cal Y}\cup\{\epsilon\}$ where $\epsilon$ (called blank) denotes the null output. ${\cal B}$ is a mapping that removes consecutive duplicates and blanks (in this order) such that ${\cal B}({\bf a})={\bf y}$. We compute the draft CTC hypothesis as $\hat{{\bf y}} = {\cal B}(\hat{{\bf a}})$ with

\begin{equation}
\hat{{\bf a}}=(\hat{a}_1,\ldots,\hat{a}_T)=(\argmax_{a\in\overline{{\cal Y}}}p(a|h_1),\ldots,\argmax_{a\in\overline{{\cal Y}}}p(a|h_T))
\end{equation}
~~\\
denoting the greedy alignment path. The draft hypothesis is accepted as final if all the frame-level entropies of the CTC output distribution are less than a threshold i.e.

\begin{equation}
\sum_{a\in\overline{\cal Y}}-p(a|h_t)\log p(a|h_t) < \tau_{CTC}\qquad\forall t\in\{1,\ldots,T\}
\label{eq:CTC-verif}
\end{equation}

\subsection{SLM verification of the CTC hypothesis}
\label{subsec:SLM-verif}
In comparison to~(\ref{ll}), SLMs model the conditional distribution of the output sequence ${\bf y}$ given the input sequence ${\bf x}$ differently:
\begin{equation}
p_{SLM}({\bf y}|{\bf x})=\prod_{i=1}^U p(y_i|{\bf y}_{<i},{\bf x})=\prod_{i=1}^U p(y_i|{\bf y}_{<i},{\bf g})
\end{equation}
~~\\
where ${\bf g}=(g_1,\ldots,g_{T'})=Adapter({\bf h})$, $T'\le T$, is a sequence of LLM-compatible acoustic embeddings computed by a speech modality adapter that takes as input the embedding sequence ${\bf h}$ computed by the CTC encoder. The draft CTC hypothesis $\hat{\bf y}$ is accepted during the SLM verification pass if all the token likelihoods under the SLM output distribution are greater than a threshold:

\begin{equation}
p(\hat{y}_i|\hat{{\bf y}}_{<i},{\bf g}) > \tau_{SLM}\qquad\forall i\in\{1,\ldots,|\hat{\bf y}|\}
\label{eq:SLM-verif}
\end{equation}

Note that, because of the causal attention mask of the text LLM,~(\ref{eq:SLM-verif}) can be computed in parallel for all tokens with a single forward pass through the LLM. The reason why we use entropy in~(\ref{eq:CTC-verif}) and likelihood in~(\ref{eq:SLM-verif}) is because the specific hypothesis $\hat{{\bf y}}$ needs to be verified for the latter.  

\subsection{Auto-regressive fallback pass with CTC prefix}
\label{subsec:SLM-AR}
If~(\ref{eq:SLM-verif}) does not hold, we find the longest verified CTC prefix $\hat{{\bf y}}_{<j}$ where $j=\min\{i\in\{1,\ldots,|\hat{\bf y}|\}~s.t.~p(\hat{y}_i|\hat{{\bf y}}_{<i},{\bf g}) \le \tau_{SLM}\}$ and simply continue auto-regressive generation from there:

\begin{equation}
\overline{\bf y}=\argmax_{{\bf y}\in{\cal Y}^*}p_{SLM}({\bf y}|\hat{{\bf y}}_{<j},{\bf x})
\end{equation}
~~\\
In summary, the final hypothesis that is output by the self-speculative decoding (short SSD) process is

\begin{equation}
{\bf y}_f=\left\{
\begin{array}{ll}
\hat{\bf y}, &\mathrm{if}~(\ref{eq:CTC-verif})~\mathrm{or}~(\ref{eq:SLM-verif})\\
(\hat{{\bf y}}_{<j},\overline{\bf y}),&\mathrm{otherwise}
\end{array}
\right.
\end{equation}

 \begin{table}[!t]                                                                                                                                                                                                           
    \centering                                                                                                                                                                                                                
    \caption{Comparing full AR and self-speculative decoding for granite-speech-4.0-1b. High accuracy: $\tau_{\text{CTC}}\!=\!0.7, \tau_{\text{SLM}}\!=\!0.2$. High RTFx: $\tau_{\text{CTC}}\!=\!3.0,                         
  \tau_{\text{SLM}}\!=\!0.1$.  \%C/\%L = CTC/LLM acceptance rates at each verification stage. RTFx are measured on 1 H100. $^{*}p<0.0001$, $^{\dagger}p<0.05$ (Statistical significance was assessed using two-proportion z-test).}                 
    \label{tab:results}                                                                                                                                                                                                       
    \footnotesize                                                                                                                                                                                                             
    \setlength{\tabcolsep}{3pt}
    \begin{tabular}{@{}l rr rr@{\hspace{3pt}}r@{\hspace{3pt}}r rr@{}}
    \toprule                                                                                                                                                                                                                  
    & \multicolumn{2}{c}{\textbf{Full AR}} & \multicolumn{4}{c}{\textbf{High Accuracy}} & \multicolumn{2}{c}{\textbf{High RTFx}} \\                                                                                           
    \cmidrule(lr){2-3} \cmidrule(lr){4-7} \cmidrule(lr){8-9}                                                                                                                                                                  
    \textbf{Test set} & WER & RTFx & WER & RTFx & \%C & \%L & WER & RTFx \\                                                                                                                                                   
    \midrule                                                                                                                                                                                                                  
    \multicolumn{9}{@{}l}{\textit{English (Open ASR)}} \\                                                                                                                                                                     
    AMI\_IHM    & 8.65 & 455 & 8.31 & 621 & 39 & 39 & 9.40 & 1716 \\                                                                                                                                                          
    Earnings22  & 9.19 & 391 & 8.96 & 278 &  6 & 19 & 11.44& 1982 \\                                                                                                                                                          
    GigaSpeech  &10.19 & 511 & 9.95 & 514 & 10 & 51 & 10.67& 2503 \\                                                                                                                                                          
    LS Clean    & 1.37 & 510 & 1.37 & 712 & 17 & 73 & 1.63 & 2160 \\                                                                                                                                                          
    LS Other    & 3.02 & 476 & 2.88 & 667 & 18 & 61 & 3.77 & 2108 \\                                                                                                                                                          
    SPGISpeech  & 3.97 & 666 & 3.80 & 592 &  6 & 44 & 4.48 & 2736 \\                                                                                                                                                          
    Tedlium     & 3.38 & 358 & 3.35 & 444 &  9 & 60 & 3.91 & 1748 \\                                                                                                                                                          
    VoxPopuli   & 6.21 & 304 & 5.99 & 365 & 12 & 51 & 7.18 & 2035 \\                                                                                                                                                          
    \rowcolor{gray!15}                                                                                                                                                                                                        
    \textbf{Avg.} & \textbf{5.75} & \textbf{564} & \textbf{5.58}\rlap{$^{*}$} & \textbf{548} & 14 & 47 & \textbf{6.56} & \textbf{2491} \\                                                                                          
    \midrule                                                                                                                                                                                                                  
    \multicolumn{9}{@{}l}{\textit{Multilingual (MLS)}} \\                                                                                                                                                                     
    MLS En      & 4.74 & 702 & 4.72 & 772 &  1 & 42 & 5.75 & 2983 \\                                                                                                                                                          
    MLS De      & 4.59 & 663 & 4.42 & 755 &  3 & 48 & 4.91 & 2755 \\                                                                                                                                                          
    MLS Es      & 3.28 & 627 & 3.22 & 779 &  3 & 50 & 3.66 & 2490 \\                                                                                                                                                          
    MLS Fr      & 4.51 & 634 & 4.48 & 658 &  1 & 39 & 5.52 & 2834 \\                                                                                                                                                          
    MLS Pt      &10.25 & 521 & 9.84 & 531 &  8 & 10 & 8.83 & 2701 \\                                                                                                                                                          
    \rowcolor{gray!15}                                                                                                                                                                                                        
    \textbf{Avg.} & \textbf{5.47} & \textbf{629} & \textbf{5.33}\rlap{$^{\dagger}$} & \textbf{699} & 3 & 38 & \textbf{5.73} & \textbf{2753} \\                                                                                             
    \midrule                                                                                                                                                                                                                  
    \multicolumn{9}{@{}l}{\textit{Multilingual (CommonVoice)}} \\                                                                                                                                                             
    CV En       & 6.52 & 954 & 6.56 & 716 & 28 & 31 & 9.30 & 2446 \\                                                                                                                                                          
    CV De       & 4.79 & 872 & 4.71 & 998 &  0 & 59 & 6.34 & 2309 \\                                                                                                                                                          
    CV Es       & 4.04 & 752 & 3.98 & 890 &  1 & 59 & 5.42 & 2526 \\                                                                                                                                                          
    CV Fr       & 7.15 & 674 & 7.10 & 910 &  0 & 46 & 10.69& 2341 \\                                                                                                                                                          
    CV Pt       & 2.82 & 866 & 2.80 &1066 &  5 & 69 & 3.83 & 2344 \\                                                                                                                                                          
    \rowcolor{gray!15}                                                                                                                                                                                                        
    \textbf{Avg.} & \textbf{5.06} & \textbf{824} & \textbf{5.03} & \textbf{916} & 7 & 53 & \textbf{7.11} & \textbf{2393} \\                                                                                          
    \bottomrule                                                                                                                                                                                                               
    \end{tabular}                                                                                                                                                                                                             
  \end{table}
\begin{figure}
\centering
\includegraphics[width=8cm]{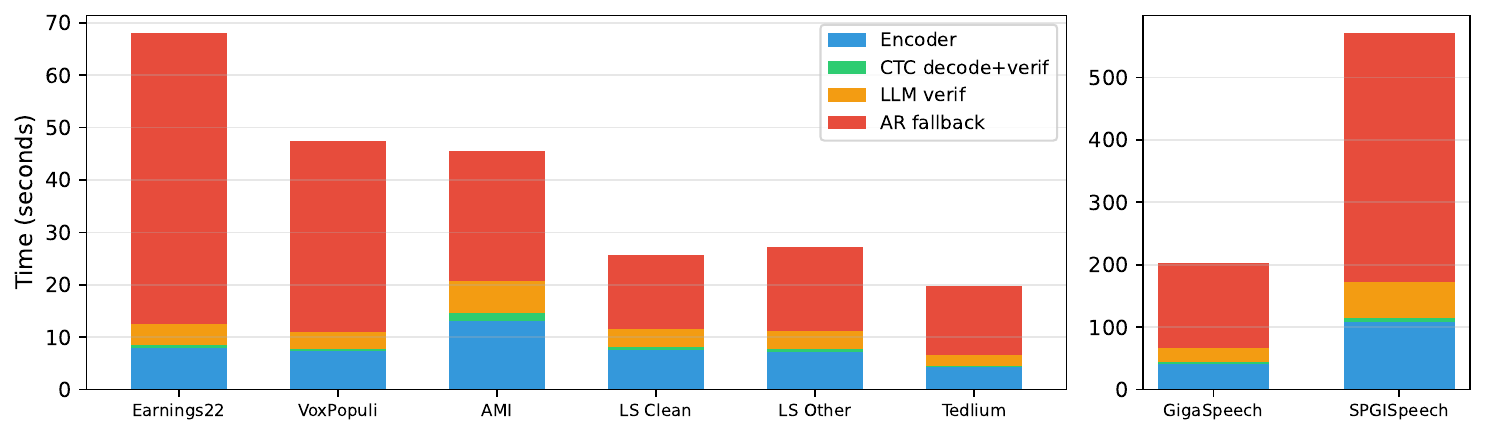}
\caption{Run times and breakdown for the different passes in the high accuracy SSD regime for granite-speech-4.0-1b.}
\label{fig:timing}
\end{figure}                                                        

\section{Experiments and results}
\label{sec:experiments}
We selected Granite Speech~\cite{saon2025granite} as the SLM for our experiments because it satisfies the constraint of having a frozen acoustic encoder trained with CTC and has shown excellent performance on the Open ASR leaderboard~\cite{srivastav2025open}. We report results for the existing granite-speech-3.3-2b and granite-speech-3.3-8b models as well as a newly trained SLM on top of a 1B parameter granite-4.0-nano LLM~\cite{granite2025}. We discuss the architecture and training specifics for the latter in terms of training and evaluation data, encoder architecture and CTC training, followed by projector architecture and training with LLM finetuning.


\subsection{Data, architecture and training specifics}
Our model is trained on 90,000 hours of data from major publicly available multilingual ASR datasets in English, French, German, Spanish, Portuguese and Japanese as well as bidirectional synthetic translations of CommonVoice to and from English to support the speech translation task. Concretely, our CTC encoder is trained on Multilingual LibriSpeech~\cite{pratap2020mls}, CommonVoice 17.0~\cite{ardila2019common}, LibriSpeech~\cite{panayotov2015librispeech}, Voxpopuli~\cite{wang2021voxpopuli}, AMI~\cite{kraaij2005ami}, YODAS~\cite{li2023yodas}, Earnings-22~\cite{delrio2022earnings22} (train partition cf.~\cite{gandhi2022esbbenchmarkmultidomainendtoend}), Switchboard~\cite{godfrey1992switchboard}, CallHome, Fisher~\cite{cieri2004fisher}, Voicemail~\cite{padmanabhan2002automatic} and ReazonSpeech~\cite{yin2023reazonspeech}. Projector training and LLM finetuning use the previously mentioned corpora plus 21,000 hours of synthetic translation data using Phi-4~\cite{abdin2024phi}, MADLAD~\cite{kudugunta2023madlad} and Granite 3~\cite{granite2024granite}  translations. We also report ASR results on Gigaspeech~\cite{chen2021gigaspeech}, SPGI Speech~\cite{o2021spgispeech} and TED LIUM~\cite{rousseau2012ted} which were excluded from our training data because of non-commercial licenses.

The 440M parameter encoder consists of 16 conformer layers of hidden dimension 1024 with 4-second block self-attention and conditioning on intermediate predictions from the middle layer. The output layer of size 348 contains the first 256 ASCII character entries to cover the European languages plus 92 Katakana phonetic characters for Japanese. It was trained with character-level CTC on all the ASR corpora specified previously for 20 epochs with a batch size of 256 using a mix of balanced and natural distribution sampling. 

We use a two-layer 37M parameter query transformer (q-former) projector that downsamples the acoustic embeddings by a factor of 5 for every window of 15 consecutive frames using 3 trainable queries per layer~\cite{tang2023salmonn}. The initial 10ms 80-dimensional log-mel frames are temporally downsampled by a factor of 10 (2x from the encoder, 5x from the projector). The projector and rank-64 LoRA adapters for the linear LLM modules are trained jointly for 3 epochs, 900k updates, with a peak learning rate of 1e-4 and a batch size of 128 utterances on all the corpora mentioned before using balanced sampling.

\subsection{Experimental details}
For all experiments, we measure WER and inverse real-time factor (RTFx = audio duration divided by processing time) on 1 H100 in a batched decoding setting using bfloat16 precision. To saturate the GPU, we sort the utterances by audio length and use adaptive batching controlled by a maximum number of tokens (set to 50K in practice) and perform separate batching for the CTC+LLM verification phases and the AR fallback phase with CPU-offloading of the acoustic embeddings for the utterances that fail verification between the passes. Additional optimizations include: (1) merging LoRA parameters and LLM weights to avoid LoRA computation overhead; (2) separate instantiation of the LLM for granite-speech to enable flash attention (FA2) since granite-speech does not support FA2 natively; (3) computing LLM logits only at CTC verification positions instead of the entire prompt that includes the text instruction and acoustic embeddings.

\begin{figure}[!t]
\centering
\includegraphics[width=8cm]{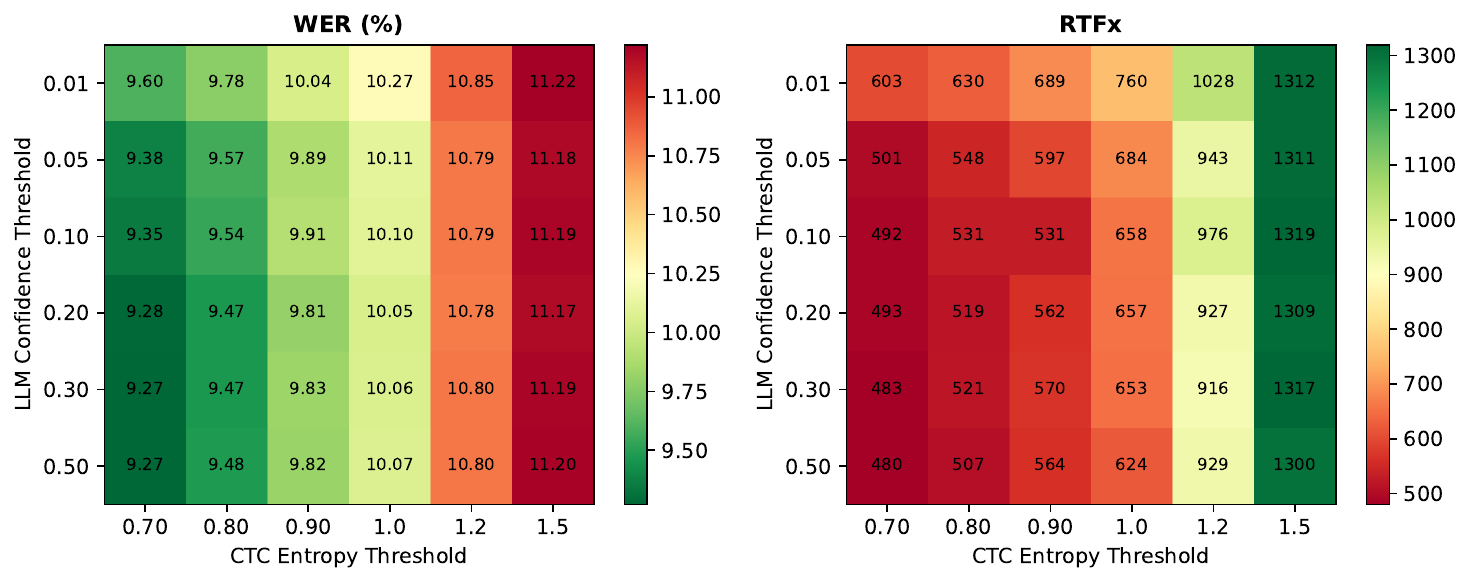}
\caption{Influence of acceptance thresholds $\tau_{CTC}$ and $\tau_{SLM}$ on word error rate and inverse real-time factor for Earnings-22.}
\label{fig:heatmap}
\end{figure}                                                        

In Figure~\ref{fig:heatmap} we study the influence of the CTC and LLM verification thresholds on WER and RTFx for the Earnings-22 corpus and conclude that $\tau_{CTC}$ has a much stronger influence on the operating point. Therefore, we fix $\tau_{SLM}=0.1$ throughout the experiments and only vary $\tau_{CTC}\in[0.7,3.0]$ except for the high accuracy settings in Table~\ref{tab:results} where $\tau_{SLM}=0.2$.  

\subsection{Verification pass ablation study}
\begin{figure}[!t]
\centering
\includegraphics[width=8cm]{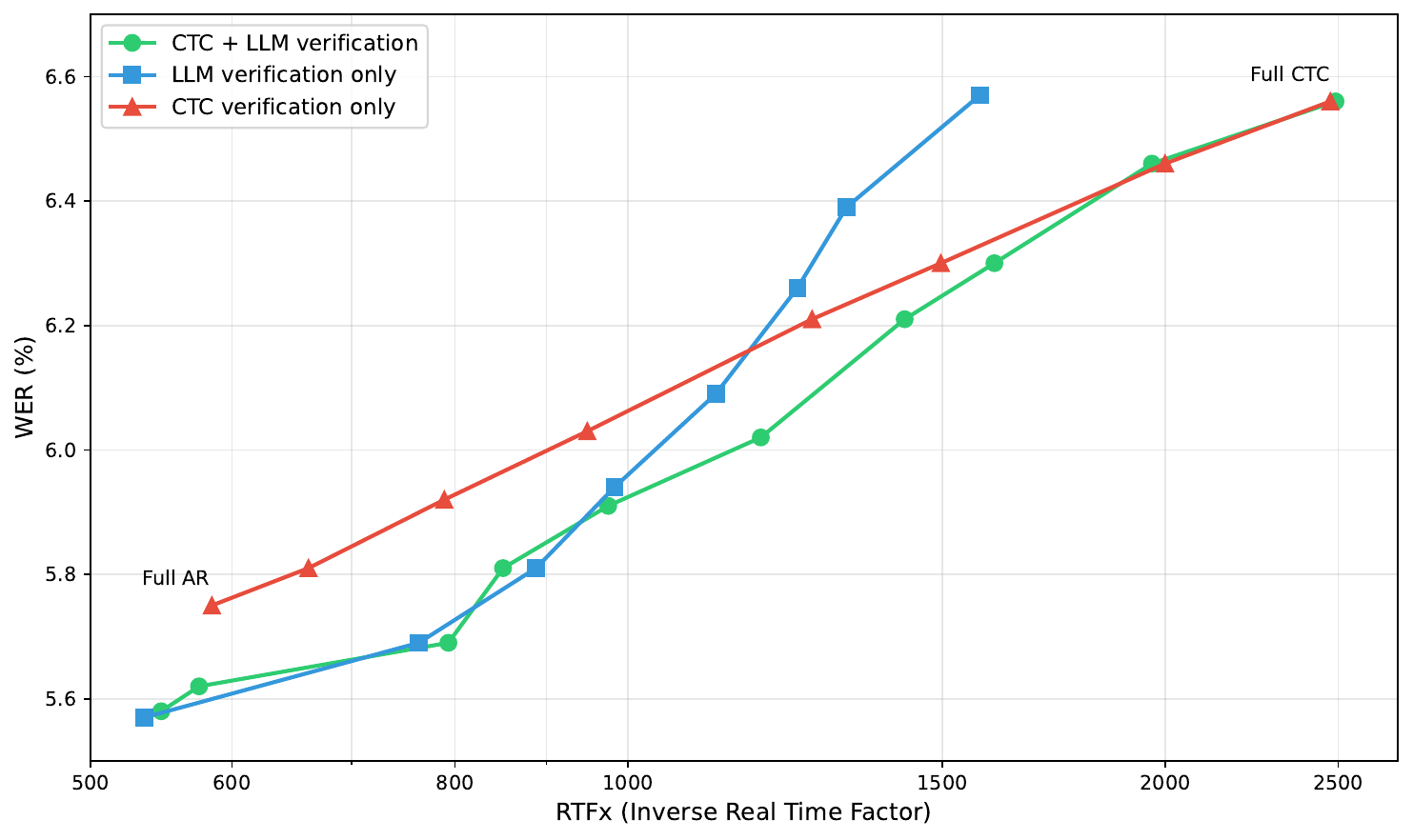}
\caption{Ablation study of verification passes: green = CTC followed by LLM; blue = LLM verification; red = CTC verification. Results are reported on the ESB/Open ASR test sets.}
\label{fig:ablation}
\end{figure}

Here we answer the question if both verification stages are necessary for optimal performance across the entire WER-RTFx spectrum. Concretely, we contrast using both CTC and LLM acceptance passes with using a single verification stage (either CTC or LLM). The results from Figure~\ref{fig:ablation} suggest that deactivating the CTC acceptance pass (i.e. directly sending the CTC hypotheses to LLM verification) has similar performance to the baseline in the high accuracy regime but has substantially lower RTFx at the high throughput end. Conversely, deactivating the LLM verification pass (i.e. using only CTC entropy acceptance and full AR fallback for rejected samples) cannot reach the lowest WER in the high accuracy regime (5.75\% vs. 5.58\%) but does match the baseline curve for high RTFx. The proposed method using both verification steps dominates across most of the range and achieves the best Pareto frontier.

\subsection{LLM verification improves accuracy over full AR}
As can be seen from the results in Table~\ref{tab:results} and Figure~\ref{fig:ablation}, the LLM verification stage of the CTC hypotheses reduces WER across all corpora compared to full auto-regressive decoding. We surmise that this is due to a system combination effect between CTC and SLM which make complementary recognition errors. In Table~\ref{tab:examples} we show some examples where the LLM-verified CTC hypotheses are more accurate than those obtained with full AR decoding. It is apparent that the AR outputs tend to be more fluent but not always faithful to the acoustics which is a well-known problem for encoder-decoder models such as Whisper called language model bias~\cite{koenecke2024careless}.

\begin{figure}[!t]
\centering
\includegraphics[width=8cm]{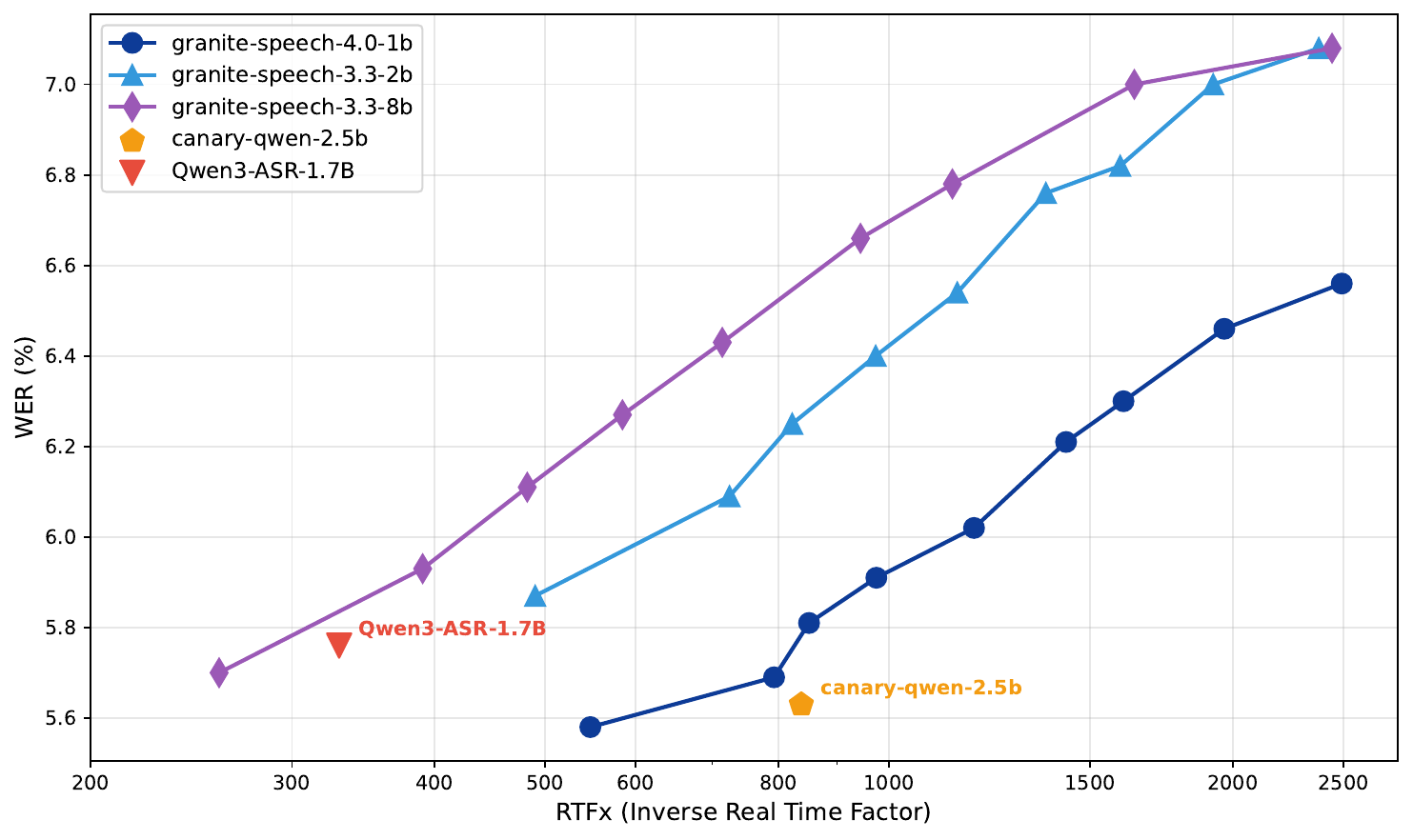}
\caption{WER and RTFx comparison between granite-speech and leading SLMs on the Open ASR test sets (all on 1 H100).}
\label{fig:comparison}
\end{figure}     

\begin{table}
\centering
\caption{Examples where LLM-verified CTC hypotheses are more accurate than full AR decoding.
Errors are shown in \textbf{bold}.}
\label{tab:examples}
\tiny
\begin{tabular}{l}
\toprule
\textbf{AMI} \\
\midrule
\texttt{REF: so if you can work on it that and then ~~like no actually not} \\
\texttt{CTC: so if you can work on it that and then ~~like no actually not} \\
\texttt{AR:~ so if you can work on \textbf{**} that and then \textbf{i} like no actually not} \\
\midrule
\textbf{LibriSpeech test-other} \\
\midrule
\texttt{REF: and i~tanquam ~sponsus} \\
\texttt{CTC: and i~tanquam ~spons\textbf{i}s} \\
\texttt{AR:~ and i~tan\textbf{t}quam spons\textbf{i}s} \\
\midrule
\texttt{REF: what chucked him off yonder} \\
\texttt{CTC: what \textbf{tr}ucked him off yonder} \\
\texttt{AR:~ what \textbf{tr}ucked \textbf{em}~~off yonder} \\
\midrule
\textbf{Voxpopuli} \\
\midrule
\texttt{REF: business~~ is ~demanding workers are demanding we are demanding do more} \\
\texttt{CTC: business~~ is ~demanding workers are demanding we are demanding do more}\\
\texttt{AR:~ business\textbf{es} \textbf{are} demanding workers are demanding we are demanding do more}\\
\midrule
\texttt{REF: i declare ~~particular interest on this matter} \\
\texttt{CTC: i declare ~~particular interest \textbf{in} this matter} \\
\texttt{AR:~ i declare \textbf{a} particular interest \textbf{in} this matter} \\

\bottomrule
\end{tabular}
\end{table}

\subsection{Detailed results}
In Table~\ref{tab:results}, we report WER and RTFx results for granite-speech-4.0-1b on the ESB/Open ASR English test sets as well as on two multilingual corpora (CommonVoice 17.0 and MLS). We observe that, in the high accuracy regime, SSD outperforms full AR across all corpora with no loss in throughput. The improvement in accuracy stems from a low CTC acceptance rate (column \%C) and a high LLM acceptance rate of 40-50\% of CTC hypotheses (column \%L). In the high RTFx regime, we accelerate inference by a factor of 4.4 on Open ASR with a 12\% degradation in WER by having close to 100\% CTC gating. As for runtime, Figure~\ref{fig:timing} shows that the encoder and fallback AR stages are the most time-consuming in the high accuracy regime.

Finally, in Figure~\ref{fig:comparison}, we compare three granite-speech models (3.3-2b, 3.3-8b and 4.0-1b) with SSD against the top 2 closest competitors (canary-qwen-2.5b~\cite{chen2024salm,sekoyan2025canary} and qwen3-asr-1.7b~\cite{shi2026qwen3asrtechnicalreport}). To ensure fair RTFx comparisons, we have run the Open ASR benchmark for the latter ourselves using the official HuggingFace evaluation harnesses provided for those models. 

\section{Conclusion and future work}
\label{sec:conclusion}
We have shown that both accuracy and inference speed of speech-aware LLMs can be improved by using fast CTC encoder drafts computed non-autoregressively. Accuracy is improved by LLM verification of acoustically-grounded CTC hypotheses which can mitigate errors due to language modeling bias. Throughput is improved by accepting high-confidence CTC drafts. Importantly, this is achieved without retraining the SLM or having to provide a separate draft model but by simply using the CTC head of the encoder for rapid transcription and gating and the LLM for verification and fallback AR decoding. 
Future work will look at training the encoder jointly with the LLM specifically for speculation (high LLM acceptance rate while maintaing accuracy) and at applying some of these ideas to reduce latency for real-time conversational applications.

\section{Generative AI Use Disclosure}
The code used in the experiments was written with help from a coding assistant (Claude by Anthropic). The experiments were run manually and results were manually verified. Generative AI was also used in the formatting of tables and in figure design. The rest of the paper was manually written with no AI proofreading or editing. The authors assume full responsibility and accountability for the content of this submission.

\bibliographystyle{IEEEtran}
\bibliography{mybib}

\end{document}